\documentclass[amsmath,aps,showpacs,a4paper,10pt]{revtex4}
 \usepackage{epsf}
 \usepackage{graphicx}    % Include figure files

\begin{document}

 \newcommand{\be}[1]{\begin{equation}\label{#1}}
 \newcommand{\ee}{\end{equation}}
 \newcommand{\bea}{\begin{eqnarray}}
 \newcommand{\eea}{\end{eqnarray}}
 \def\disp{\displaystyle}

 \def\gsim{ \lower .75ex \hbox{$\sim$} \llap{\raise .27ex \hbox{$>$}} }
 \def\lsim{ \lower .75ex \hbox{$\sim$} \llap{\raise .27ex \hbox{$<$}} }

 \begin{titlepage}

 \begin{flushright}
 arXiv:1004.0492
 \end{flushright}

 \title{\Large \bf Revisiting the Cosmological Constraints on
 the~Interacting~Dark Energy Models}

 \author{Hao~Wei\,}
 \email[\,email address:\ ]{haowei@bit.edu.cn}
 \affiliation{Department of Physics, Beijing Institute
 of Technology, Beijing 100081, China}

 \begin{abstract}\vspace{1cm}
 \centerline{\bf ABSTRACT}\vspace{2mm}
 In this work, we consider the cosmological constraints on
 the interacting dark energy models. We generalize the models
 considered previously by Guo {\it et al.}~\cite{r15}, Costa
 and Alcaniz~\cite{r16}, and try to discuss two general
 types of models: type~I models are characterized by
 $\rho_{_X}/\rho_m=f(a)$ and $f(a)$ can be any function of
 scale factor $a$, whereas type~II models are characterized
 by $\rho_m=\rho_{m0}\,a^{-3+\epsilon(a)}$ and $\epsilon(a)$
 can be any function of $a$. We obtain the cosmological
 constraints on the type~I and~II models with power-law,
 CPL-like, logarithmic $f(a)$ and $\epsilon(a)$ by using the
 latest observational data.
 \end{abstract}


 \maketitle

 \end{titlepage}

 \renewcommand{\baselinestretch}{1.4}

%============================= section 1 ===================================

\section{Introduction}\label{sec1}

The dark energy has been one of the most active fields in
 modern cosmology since the discovery of the accelerated
 expansion of our universe (see e.g.~\cite{r1} for reviews).
 Among the conundrums in the dark energy cosmology, the
 so-called cosmological coincidence problem is the most
 familiar one. This problem is asking why are we living in
 an epoch in which the densities of dark energy and matter
 are comparable? Since their densities scale differently
 with the expansion of our universe, there should be some
 fine-tunings. To alleviate the cosmological coincidence
 problem, it is natural to consider the possible interaction
 between dark energy and dark matter in the literature (see
 e.g.~\cite{r2,r3,r4,r5,r6,r7,r8,r9,r31,r32}). In fact, since
 the nature of both dark energy and dark matter are still
 unknown, there is no physical argument to exclude the possible
 interaction between them. On the contrary, some observational
 evidences of this interaction have been found recently. For
 example, in a series of papers by Bertolami
 {\it et al.}~\cite{r10}, they shown that the Abell Cluster A586
 exhibits evidence of the interaction between dark energy and
 dark matter, and they argued that this interaction might imply
 a violation of the equivalence principle. On the other hand,
 in~\cite{r11}, Abdalla {\it et al.} found the signature of
 interaction between dark energy and dark matter by using
 optical, X-ray and weak lensing data from 33 relaxed galaxy
 clusters. Therefore, it is reasonable to consider the
 interaction between dark energy and dark matter in cosmology.

We consider a flat Friedmann-Robertson-Walker (FRW) universe.
 In the literature, it is usual to assume that dark energy and
 dark matter interact through a coupling term $Q$, according to
 \bea
 &&\dot{\rho}_m+3H\rho_m=Q,\label{eq1}\\
 &&\dot{\rho}_{_X}+3H\rho_{_X}(1+w_{_X})=-Q,\label{eq2}
 \eea
 where $\rho_m$ and $\rho_{_X}$ are densities of dark matter
 and dark energy (we assume that the baryon component can be
 ignored); $w_{_X}$ is the equation-of-state parameter (EoS)
 of dark energy and it is assumed to be a constant; a dot
 denotes the derivative with respect to cosmic time $t$;
 $H\equiv\dot{a}/a$ is the Hubble parameter; $a=(1+z)^{-1}$
 is the scale factor (we have set $a_0=1$; the subscript ``0''
 indicates the present value of corresponding quantity; $z$ is
 the redshift). Notice that Eqs.~(\ref{eq1}) and (\ref{eq2})
 preserve the total energy conservation equation
 \be{eq3}
 \dot{\rho}_{tot}+3H\rho_{tot}(1+w_{\rm eff})=0,
 \ee
 where $\rho_{tot}=\rho_{_X}+\rho_m$ is the total energy;
 $w_{\rm eff}$ is the total (effective) EoS. Since there is
 no natural guidance from fundamental physics on the coupling
 term $Q$, one can only discuss it to a phenomenal level. The
 most familiar coupling terms extensively considered in the
 literature are $Q=\alpha\kappa\rho_m\dot{\phi}$,
 $Q=3\beta H\rho_{tot}$, and $Q=3\eta H\rho_m$. The first one
 arises from, for instance, string theory or scalar-tensor
 theory (including Brans-Dicke theory)~\cite{r4,r5,r6}. The
 other two are phenomenally proposed to alleviate the
 coincidence problem in the other dark energy
 models~\cite{r7,r8,r9}. In the usual approach, one should
 priorly write down the coupling term $Q$, and then obtain
 the evolutions of $\rho_m$ and $\rho_{_X}$ from
 Eqs.~(\ref{eq1}) and (\ref{eq2}), respectively. In fact, this
 is the common way to study the interacting dark energy models
 in the literature.

However, there is also an alternative way in the
 literature~\cite{r12,r13,r14,r15,r16}. One can reverse the
 logic mentioned above. Due to the interaction $Q$, the
 evolutions of $\rho_m$ and $\rho_{_X}$ should deviate from
 the ones without interaction, i.e., $\rho_m\propto a^{-3}$
 and $\rho_{_X}\propto a^{-3(1+w_{_X})}$, respectively.
 If the deviated evolutions of $\rho_m$ and/or $\rho_{_X}$
 are given, one can find the corresponding interaction $Q$
 from Eqs.~(\ref{eq1}) and (\ref{eq2}). Naively, the simplest
 example has been considered by Wang and Meng~\cite{r12},
 namely
 \be{eq4}
 \rho_m=\rho_{m0}\,a^{-3+\epsilon},
 \ee
 where $\epsilon$ is a constant which measures the deviation
 from the normal $\rho_m\propto a^{-3}$. Substituting into
 Eq.~(\ref{eq1}), it is easy to find the corresponding
 interaction $Q=\epsilon H\rho_m$~\cite{r9,r12,r13,r17}.
 Alternatively, one can consider another type of interacting
 dark energy model which is characterized by~\cite{r14,r15}
 \be{eq5}
 \frac{\rho_{_X}}{\rho_m}=\frac{\rho_{_{X}0}}{\rho_{m0}}\,a^\xi,
 \ee
 where $\xi$ is a constant which measures the severity of the
 coincidence problem. From Eqs.~~(\ref{eq1}), (\ref{eq2})
 and~(\ref{eq5}), one can find that the corresponding
 interaction is given by~\cite{r15}
 \be{eq6}
 Q=-H\rho_m\Omega_X\left(\xi+3w_{_X}\right)
 =-H\rho_{_X}\Omega_m\left(\xi+3w_{_X}\right),
 \ee
 where $\Omega_i\equiv 8\pi G\rho_i/(3H^2)$ for $i=m$ and $X$,
 which are the fractional energy densities of dark matter and
 dark energy, respectively. In fact, Guo {\it et al.}~\cite{r15}
 considered the cosmological constraints on the interacting
 dark energy model characterized by Eq.~(\ref{eq5}) with the 71
 SNLS Type~Ia supernovae (SNIa) dataset, the shift parameter $R$
 from the Wilkinson Microwave Anisotropy Probe 3-year (WMAP3)
 data, and the distance parameter $A$ of the measurement of the
 BAO peak in the distribution of SDSS luminous red galaxies.
 On the other hand, the interacting dark energy model characterized
 by Eq.~(\ref{eq4}) has been extended in~\cite{r16}. It is more
 realistic that $\epsilon$ is a function of time. Costa and
 Alcaniz~\cite{r16} considered the interacting dark energy
 model characterized by
 \be{eq7}
 \rho_m=\rho_{m0}\,a^{-3+\epsilon(a)},
 \ee
 in which $\epsilon(a)$ was chosen to be
 \be{eq8}
 \epsilon(a)=\epsilon_0\,a^{\epsilon_1},
 \ee
 where $\epsilon_0$ and $\epsilon_1$ are constants. They obtained
 the constraints on this model by using the 307 Union SNIa dataset,
 the CMB constraint $\Omega_{m0}h^2=0.109\pm 0.006$ from the
 Wilkinson Microwave Anisotropy Probe 5-year (WMAP5) data, and
 the distance ratio from $z_{BAO}=0.35$ to $z_{LS}=1089$
 measured by SDSS, namely $R_{BAO/LS}=0.0979\pm 0.0036$.

In the present work, we generalize the interacting dark
 energy models considered in~\cite{r15,r16}, and we call them
 type~I and II models, respectively. The type~I models are
 characterized by
 \be{eq9}
 \frac{\rho_{_X}}{\rho_m}=f(a),
 \ee
 where $f(a)$ can be any function of $a$, beyond the special
 case in Eq.~(\ref{eq5}). The type~II models are characterized
 by Eq.~(\ref{eq7}) but $\epsilon(a)$ can be any function of
 $a$, beyond the special case in Eq.~(\ref{eq8}). In the
 present work, we consider the constraints on these models by
 using the latest cosmological observations, namely, the 397
 Constitution SNIa dataset~\cite{r18}, the shift parameter $R$
 from the newly released Wilkinson Microwave Anisotropy Probe
 7-year (WMAP7) data~\cite{r19}, and the distance parameter $A$
 of the measurement of the BAO peak in the distribution of SDSS
 luminous red galaxies~\cite{r20,r21}. In the next section, we
 briefly introduce these observational data. In Sec.~\ref{sec3}
 and Sec.~\ref{sec4}, we discuss the type~I and~II models, and
 consider their cosmological constraints, respectively. A brief
 summary is given in Sec.~\ref{sec5}.

%============================= section 2 ===================================

\section{Observational data}\label{sec2}

In the present work, we will consider the latest
 cosmological observations, namely, the 397 Constitution SNIa
 dataset~\cite{r18}, the shift parameter $R$ from the newly
 released Wilkinson Microwave Anisotropy Probe 7-year (WMAP7)
 data~\cite{r19}, and the distance parameter $A$ of the
 measurement of the BAO peak in the distribution of SDSS
 luminous red galaxies~\cite{r20,r21}.

The data points of the 397 Constitution SNIa compiled
 in~\cite{r18} are given in terms of the distance modulus
 $\mu_{obs}(z_i)$. On the other hand, the theoretical
 distance modulus is defined as
 \be{eq10}
 \mu_{th}(z_i)\equiv 5\log_{10}D_L(z_i)+\mu_0\,,
 \ee
 where $\mu_0\equiv 42.38-5\log_{10}h$ and $h$ is the Hubble
 constant $H_0$ in units of $100~{\rm km/s/Mpc}$, whereas
 \be{eq11}
 D_L(z)=(1+z)\int_0^z \frac{d\tilde{z}}{E(\tilde{z};{\bf p})}\,,
 \ee
 in which $E\equiv H/H_0$, and ${\bf p}$ denotes the model
 parameters. Correspondingly, the $\chi^2$ from the 397
 Constitution SNIa is given by
 \be{eq12}
 \chi^2_{\mu}({\bf p})=\sum\limits_{i}\frac{\left[
 \mu_{obs}(z_i)-\mu_{th}(z_i)\right]^2}{\sigma^2(z_i)}\,,
 \ee
 where $\sigma$ is the corresponding $1\sigma$ error. The parameter
 $\mu_0$ is a nuisance parameter but it is independent of the data
 points. One can perform an uniform marginalization over $\mu_0$.
 However, there is an alternative way. Following~\cite{r22,r23}, the
 minimization with respect to $\mu_0$ can be made by expanding the
 $\chi^2_{\mu}$ of Eq.~(\ref{eq12}) with respect to $\mu_0$ as
 \be{eq13}
 \chi^2_{\mu}({\bf p})=\tilde{A}-2\mu_0\tilde{B}+\mu_0^2\tilde{C}\,,
 \ee
 where
 $$\tilde{A}({\bf p})=\sum\limits_{i}\frac{\left[\mu_{obs}(z_i)
 -\mu_{th}(z_i;\mu_0=0,{\bf p})\right]^2}
 {\sigma_{\mu_{obs}}^2(z_i)}\,,$$
 $$\tilde{B}({\bf p})=\sum\limits_{i}\frac{\mu_{obs}(z_i)
 -\mu_{th}(z_i;\mu_0=0,{\bf p})}{\sigma_{\mu_{obs}}^2(z_i)}\,,
 ~~~~~~~~~~
 \tilde{C}=\sum\limits_{i}\frac{1}{\sigma_{\mu_{obs}}^2(z_i)}\,.$$
 Eq.~(\ref{eq13}) has a minimum for
 $\mu_0=\tilde{B}/\tilde{C}$ at
 \be{eq14}
 \tilde{\chi}^2_{\mu}({\bf p})=
 \tilde{A}({\bf p})-\frac{\tilde{B}({\bf p})^2}{\tilde{C}}\,.
 \ee
 Since $\chi^2_{\mu,\,min}=\tilde{\chi}^2_{\mu,\,min}$
 obviously, we can instead minimize $\tilde{\chi}^2_{\mu}$
 which is independent of $\mu_0$.

There are some other observational data, such as
 the observations of cosmic microwave background (CMB)
 anisotropy~\cite{r19} and large-scale structure
 (LSS)~\cite{r20}. However, using the full data of CMB and LSS
 to perform a global fitting consumes a large amount of
 computation time and power. As an alternative, one can
 instead use the shift parameter $R$ from the CMB, and the
 distance parameter $A$ of the measurement of the baryon
 acoustic oscillation (BAO) peak in the distribution of SDSS
 luminous red galaxies. In the literature, the shift parameter
 $R$ and the distance parameter $A$ have been used extensively.
 It is argued that they are model-independent~\cite{r24},
 while $R$ and $A$ contain the main information of the
 observations of CMB and BAO, respectively.

As is well known, the shift parameter $R$ of the CMB is defined
 by~\cite{r24,r25}
 \be{eq15}
 R\equiv\Omega_{m0}^{1/2}\int_0^{z_\ast}
 \frac{d\tilde{z}}{E(\tilde{z})}\,,
 \ee
 where the redshift of recombination $z_\ast=1091.3$ which has
 been updated in the Wilkinson Microwave Anisotropy Probe
 7-year (WMAP7) data~\cite{r19}. The shift parameter $R$
 relates the angular diameter distance to the last scattering
 surface, the comoving size of the sound horizon at $z_\ast$
 and the angular scale of the first acoustic peak in CMB power
 spectrum of temperature fluctuations~\cite{r24,r25}. The value
 of $R$ has been updated to $1.725\pm 0.018$ from the WMAP7
 data~\cite{r19}. On the other hand, the distance parameter $A$
 of the measurement of the BAO peak in the distribution of SDSS
 luminous red galaxies~\cite{r20} is given by
 \be{eq16}
 A\equiv\Omega_{m0}^{1/2}E(z_b)^{-1/3}\left[\frac{1}{z_b}
 \int_0^{z_b}\frac{d\tilde{z}}{E(\tilde{z})}\right]^{2/3},
 \ee
 where $z_b=0.35$. In~\cite{r21}, the value of $A$ has been
 determined to be $0.469\,(n_s/0.98)^{-0.35}\pm 0.017$. Here
 the scalar spectral index $n_s$ is taken to be $0.963$, which
 has been updated from the WMAP7 data~\cite{r19}. So, the total
 $\chi^2$ is given by
 \be{eq17}
 \chi^2=\tilde{\chi}^2_{\mu}+\chi^2_{CMB}+\chi^2_{BAO}\,,
 \ee
 where $\tilde{\chi}^2_{\mu}$ is given in Eq.~(\ref{eq14}),
 $\chi^2_{CMB}=(R-R_{obs})^2/\sigma_R^2$ and
 $\chi^2_{BAO}=(A-A_{obs})^2/\sigma_A^2$. The best-fit model
 parameters are determined by minimizing the total $\chi^2$.
 As in~\cite{r26,r27}, the $68\%$ confidence level is determined by
 $\Delta\chi^2\equiv\chi^2-\chi^2_{min}\leq 1.0$, $2.3$ and
 $3.53$ for $n_p=1$, $2$ and $3$, respectively, where $n_p$ is
 the number of free model parameters. Similarly, the $95\%$
 confidence level is determined by
 $\Delta\chi^2\equiv\chi^2-\chi^2_{min}\leq 4.0$, $6.17$ and
 $8.02$ for $n_p=1$, $2$ and $3$, respectively.

%============================= section 3 ===================================

\section{Type I models}\label{sec3}

%============================= section 3.1 ===================================

\subsection{Equations}\label{sec3a}

As mentioned in Sec.~\ref{sec1}, the type~I models are
 characterized by Eq.~(\ref{eq9}), whereas $f(a)$ can be any
 function of $a$. From Eq.~(\ref{eq9}), it is easy to obtain
 \be{eq18}
 \Omega_{_X}=\frac{f}{1+f}\,,~~~~~~~
 \Omega_m=\frac{1}{1+f}\,.
 \ee
 Substituting $\rho_{_X}=\rho_m f(a)$ into Eq.~(\ref{eq2})
 and using $\dot{\rho}_m$ from Eq.~(\ref{eq1}), we can
 find that the corresponding interaction term is given by
 \be{eq19}
 Q=-H\rho_m\Omega_{_X}\left(a\frac{f^\prime}{f}+3w_{_X}\right)
 =-H\rho_{_X}\Omega_m\left(a\frac{f^\prime}{f}+3w_{_X}\right),
 \ee
 where a prime denotes the derivative with respect to $a$.
 Obviously, if $f(a)\propto a^\xi$, Eq.~(\ref{eq19}) reduces
 to Eq.~(\ref{eq6}) which has been obtained in~\cite{r15}.
 On the other hand, one can recast Eq.~(\ref{eq3}) as
 \be{eq20}
 \frac{d\ln\rho_{tot}}{d\ln a}=-3\left(1+w_{\rm eff}\right)
 =-3\left(1+\Omega_{_X}w_{_X}\right).
 \ee
 Using Eq.~(\ref{eq18}), we can integrate Eq.~(\ref{eq20}) to
 obtain
 \be{eq21}
 \rho_{tot}=a^{-3}\exp\left(-\int\frac{3w_{_X}f}{1+f}\,d\ln a
 \right)\cdot const.\,,
 \ee
 where $const.$ is an integral constant, which can be determined by
 requiring the condition
 $\rho_{tot}(a=1)=\rho_{tot,0}=3H_0^2/(8\pi G)$. Once $\rho_{tot}$
 is on hand, we can readily find the corresponding
 $E\equiv H/H_0$ from Friedmann equation, and then fit it to
 the observational data. Correspondingly,
 $\rho_{_X}=\Omega_{_X}\rho_{tot}$ and $\rho_m=\Omega_m\rho_{tot}$
 are also available from Eqs.~(\ref{eq18}) and (\ref{eq21}).
 Finally, it is worth noting that by definition~(\ref{eq9}),
 we have
 \be{eq22}
 f_0=f(a=1)=\frac{\rho_{_X0}}{\rho_{m0}}
 =\frac{1}{\,\Omega_{m0}}-1\,,
 \ee
 which is useful to fix one of parameters in $f(a)$.

%============================= section 3.2 ===================================

\subsection{Cosmological constraints on type I models}\label{sec3b}

\begin{center}
 \begin{figure}[tbhp]
 \centering
 \includegraphics[width=0.45\textwidth]{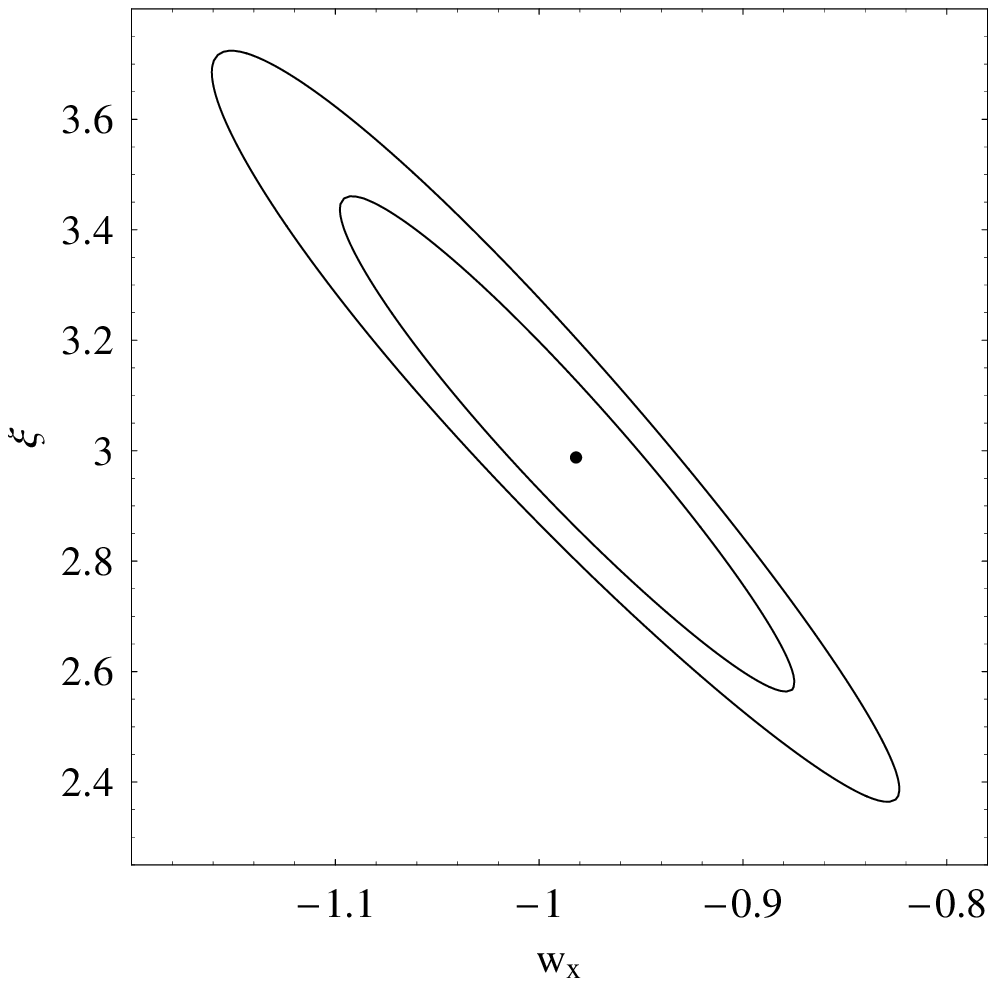}\hfill
 \includegraphics[width=0.45\textwidth]{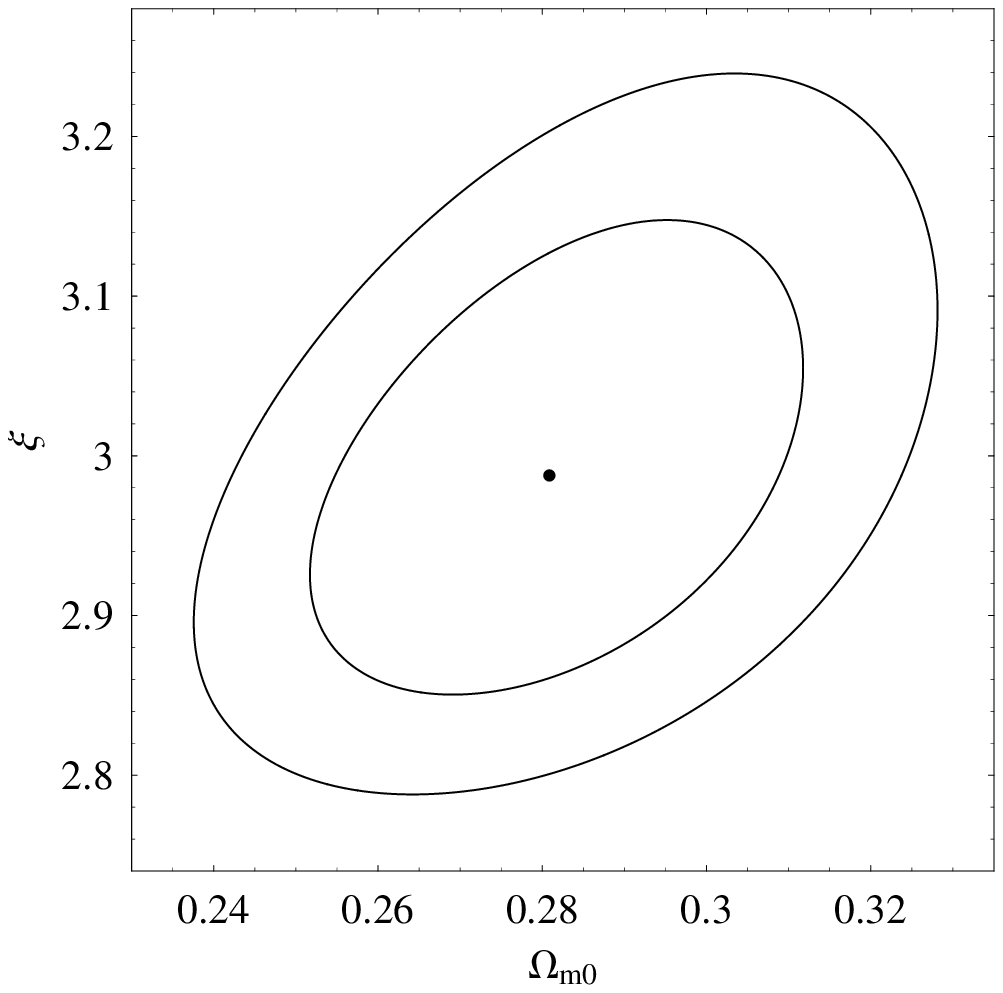}\\ \ \\
 \includegraphics[width=0.45\textwidth]{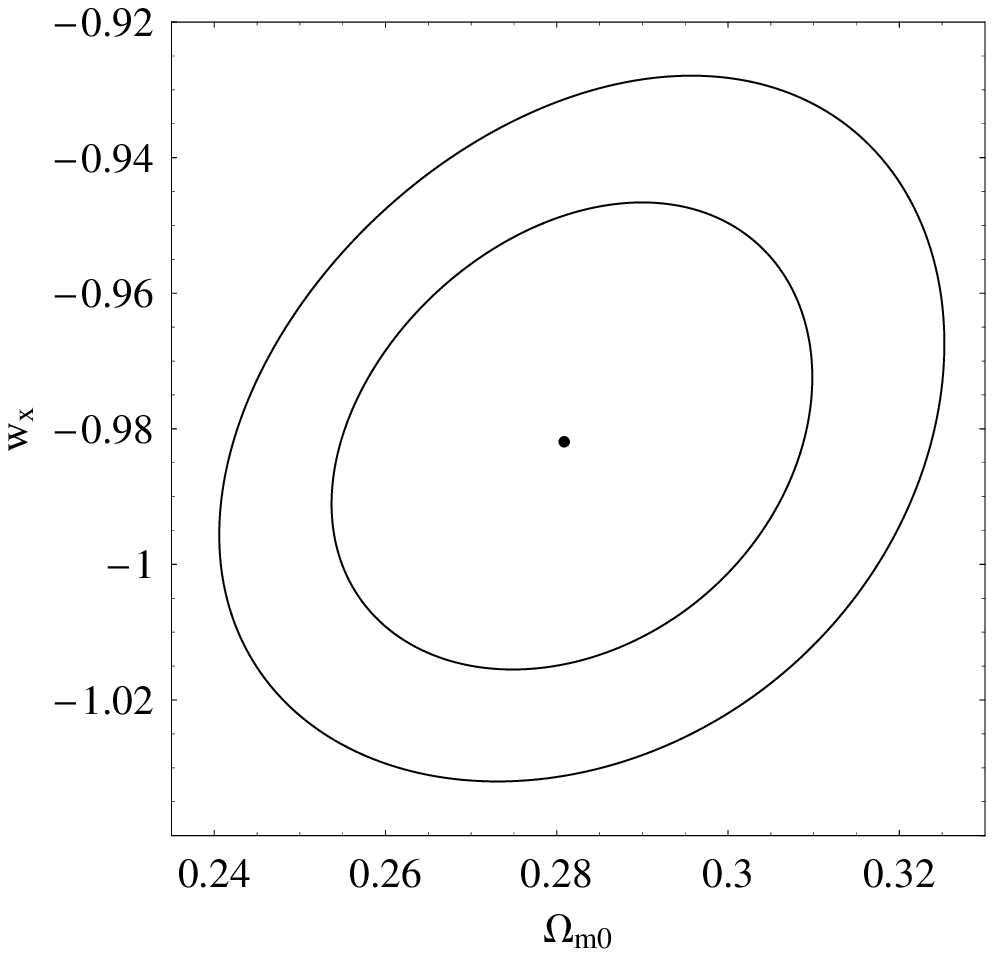}
 \caption{\label{fig1} The $68\%$ and $95\%$ confidence level
 contours in $w_{_X}-\xi$ plane, $\Omega_{m0}-\xi$ plane and
 $\Omega_{m0}-w_{_X}$ plane for the type I model characterized
 by $f(a)=f_0\,a^\xi$. The best-fit parameters are also
 indicated by the black solid points.}
 \end{figure}
 \end{center}

In this subsection, we consider the cosmological constraints
 on type I models by using the observational data given in
 Sec.~\ref{sec2}. At first, we consider the power-law case with
 \be{eq23}
 f(a)=f_0\,a^\xi,
 \ee
 where $\xi$ is a constant; $f_0$ can be determined by
 definition~(\ref{eq9}) to be the one given in Eq.~(\ref{eq22}),
 and hence it is not an independent parameter. In this case,
 there are three free model parameters, namely, $\Omega_{m0}$,
 $w_{_X}$ and~$\xi$. Although the cosmological constraints on
 the model characterized by Eq.~(\ref{eq23}) has been considered by
 Guo {\it et al.}~\cite{r15}, as mentioned in Sec.~\ref{sec1}, they
 have used the earlier observational data. Therefore, it is still
 worthwhile to consider the cosmological constraints once more
 in the present work by using the latest observational data
 mentioned in Sec.~\ref{sec2}. Substituting Eq.~(\ref{eq23})
 into Eq.~(\ref{eq21}) and requiring $\rho_{tot}(a=1)=\rho_{tot,0}$,
 we can determine the integral constant, and finally obtain
 \be{eq24}
 \rho_{tot}=\rho_{tot,0}\,a^{-3}\left[\,\Omega_{m0}+
 \left(1-\Omega_{m0}\right)a^\xi\,\right]^{-3w_{_X}/\,\xi}.
 \ee
 Substituting into Friedmann equation, we find that
 \bea
 E^2=\frac{H^2}{H_0^2}
 &=&a^{-3}\left[\,\Omega_{m0}+\left(1-\Omega_{m0}\right)a^\xi\,
 \right]^{-3w_{_X}/\,\xi}\nonumber\\
 &=&(1+z)^3 \left[\,\Omega_{m0}+\left(1-\Omega_{m0}\right)
 (1+z)^{-\xi}\,\right]^{-3w_{_X}/\,\xi}.\label{eq25}
 \eea
 By minimizing the corresponding total $\chi^2$
 in Eq.~(\ref{eq17}), we find the best-fit parameters
 $\Omega_{m0}=0.281$, $w_{_X}=-0.982$ and $\xi=2.988$,
 whereas $\chi^2_{min}=465.604$. In Fig.~\ref{fig1}, we also
 present the corresponding $68\%$ and $95\%$ confidence level
 contours in $w_{_X}-\xi$ plane, $\Omega_{m0}-\xi$ plane and
 $\Omega_{m0}-w_{_X}$ plane. It is easy to see that these
 constraints on the model characterized by $f(a)=f_0\,a^\xi$
 are much tighter than the ones obtained by
 Guo {\it et al.}~\cite{r15}, thanks to the
 latest observational data.\\

%============================= Fig. 1 =================================

%======================================================================

\vspace{-10mm}  % used here just for a more comfortable typesetting

%============================= Fig. 2 =================================

 \begin{center}
 \begin{figure}[tbhp]
 \centering
 \includegraphics[width=0.45\textwidth]{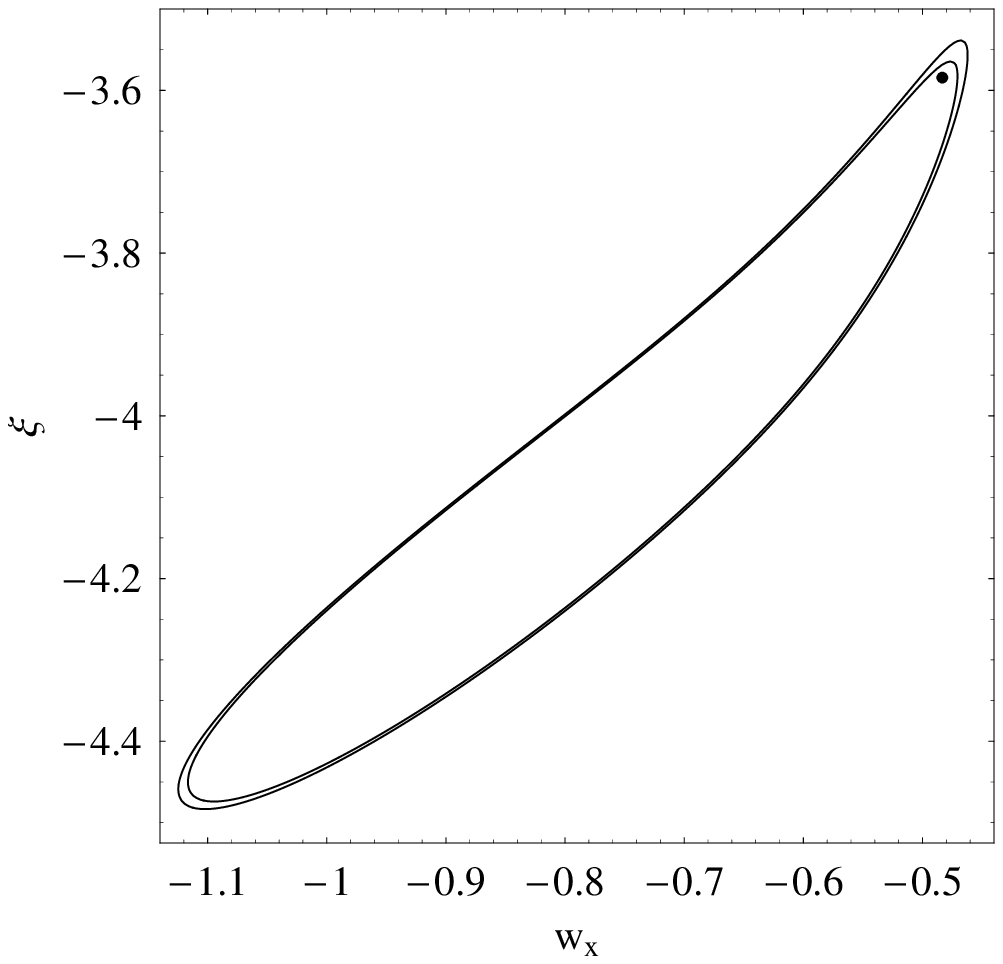}\hfill
 \includegraphics[width=0.45\textwidth]{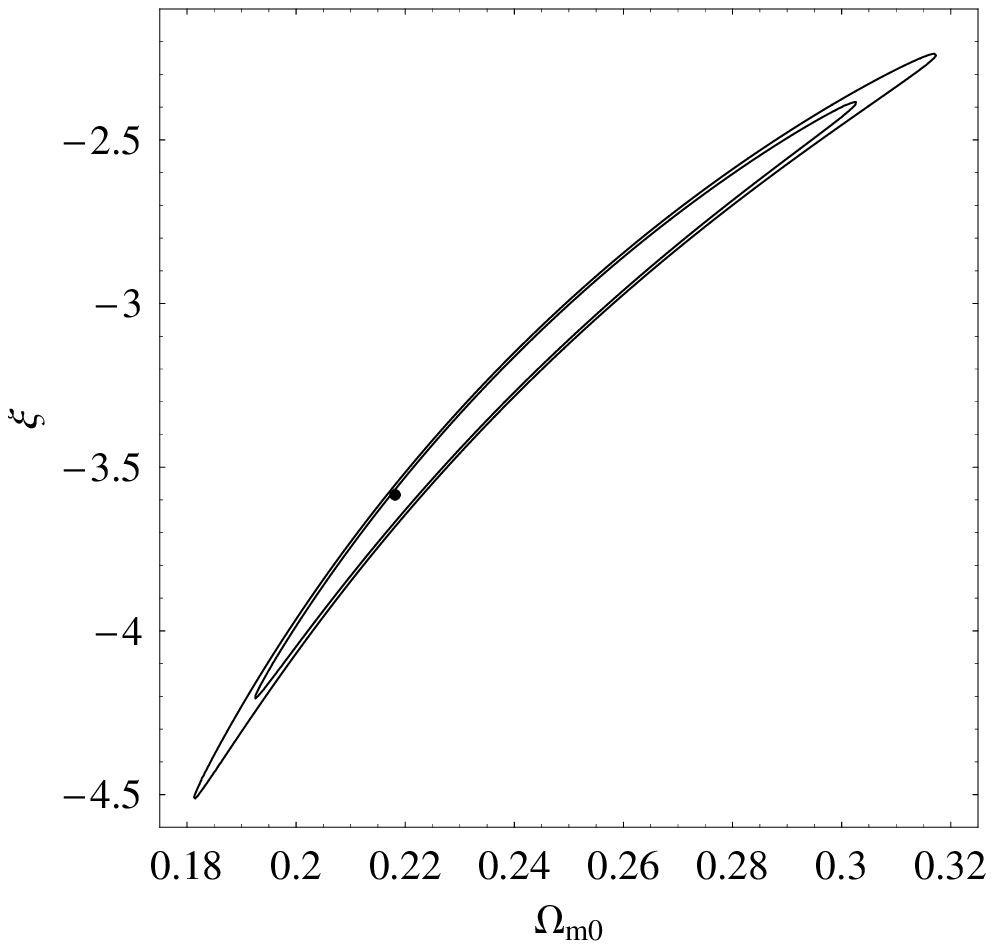}\\ \ \\
 \includegraphics[width=0.45\textwidth]{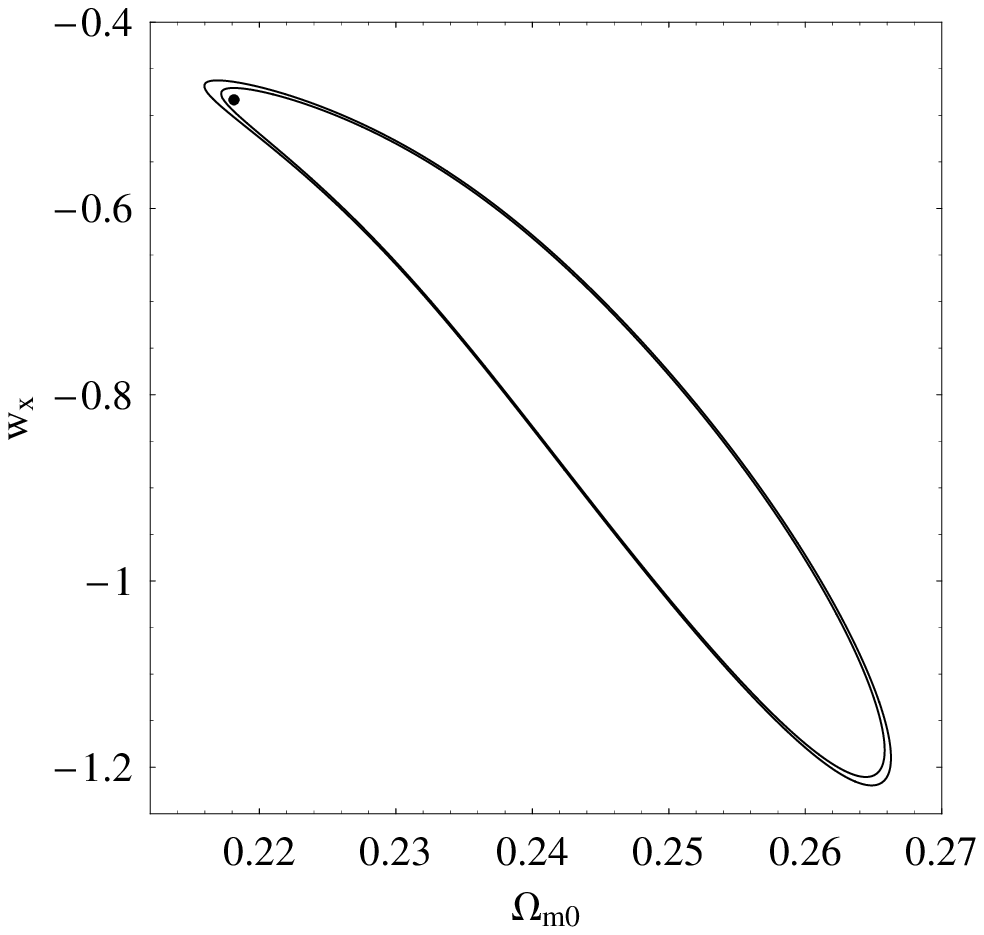}
 \caption{\label{fig2} The same as in Fig.~\ref{fig1}, except
 for the type I model characterized by $f(a)=f_0+\xi(1-a)$ with
 the condition $\xi\ge 1-\Omega_{m0}^{-1}$. See the text for
 details.}
 \end{figure}
 \end{center}

%======================================================================

\vspace{-4mm}  % used here just for a more comfortable typesetting

Next, we consider a new case with
 \be{eq26}
 f(a)=f_0+\xi(1-a),
 \ee
 which can be regarded as a linear expansion of $f(a)$ with respect
 to $a$, similar to the familiar Chevallier-Polarski-Linder (CPL)
 parameterization for EoS $w(a)=w_0+w_a(1-a)$~\cite{r28}.
 Again, $f_0$ can be determined by definition~(\ref{eq9}) to be
 the one given in Eq.~(\ref{eq22}), and hence it is not an
 independent parameter. Thus, there are three free model
 parameters, namely, $\Omega_{m0}$, $w_{_X}$ and $\xi$.
 Substituting Eq.~(\ref{eq26}) into Eq.~(\ref{eq21}) and
 requiring $\rho_{tot}(a=1)=\rho_{tot,0}$, we can determine the
 integral constant, and finally obtain
 \be{eq27}
 \rho_{tot}=\rho_{tot,0}\,a^{-3(1+w_{_X})}\left[\,
 \left(1+\xi\Omega_{m0}\right)a^{-1}-\xi\Omega_{m0}\,
 \right]^{-3w_{_X}\Omega_{m0}/\left(1+\xi\Omega_{m0}\right)}.
 \ee
 Substituting into Friedmann equation, we find that
 \bea
 E^2=\frac{H^2}{H_0^2}
 &=&a^{-3(1+w_{_X})}\left[\,\left(1+\xi\Omega_{m0}\right)a^{-1}
 -\xi\Omega_{m0}\,\right]^{-3w_{_X}\Omega_{m0}/\left(1+
 \xi\Omega_{m0}\right)}\nonumber\\
 &=&(1+z)^{3(1+w_{_X})}\left[\,\left(1+\xi\Omega_{m0}\right)(1+z)
 -\xi\Omega_{m0}\,\right]^{-3w_{_X}\Omega_{m0}/\left(1+
 \xi\Omega_{m0}\right)}.\label{eq28}
 \eea
 Noting Eq.~(\ref{eq9}) and imposing the condition $\rho_{_X}\ge 0$,
 we have $\xi\ge 1-\Omega_{m0}^{-1}$ from Eqs.~(\ref{eq26})
 and~(\ref{eq22}). Under this condition, by minimizing the
 corresponding total $\chi^2$ in Eq.~(\ref{eq17}), we find the
 best-fit parameters $\Omega_{m0}=0.218$, $w_{_X}=-0.483$ and
 $\xi=-3.584$, whereas $\chi^2_{min}=563.77$.
 In Fig.~\ref{fig2}, we also present the corresponding $68\%$
 and $95\%$ confidence level contours in $w_{_X}-\xi$ plane,
 $\Omega_{m0}-\xi$ plane and $\Omega_{m0}-w_{_X}$ plane. It is
 easy to see that the $68\%$ and $95\%$ confidence level
 contours are very close. On the other hand,
 $\chi^2_{min}=563.77$ is fairly larger than the degree of
 freedom $dof\sim 400$.

%============================= Fig. 3 =================================

 \begin{center}
 \begin{figure}[tbhp]
 \centering
 \includegraphics[width=0.45\textwidth]{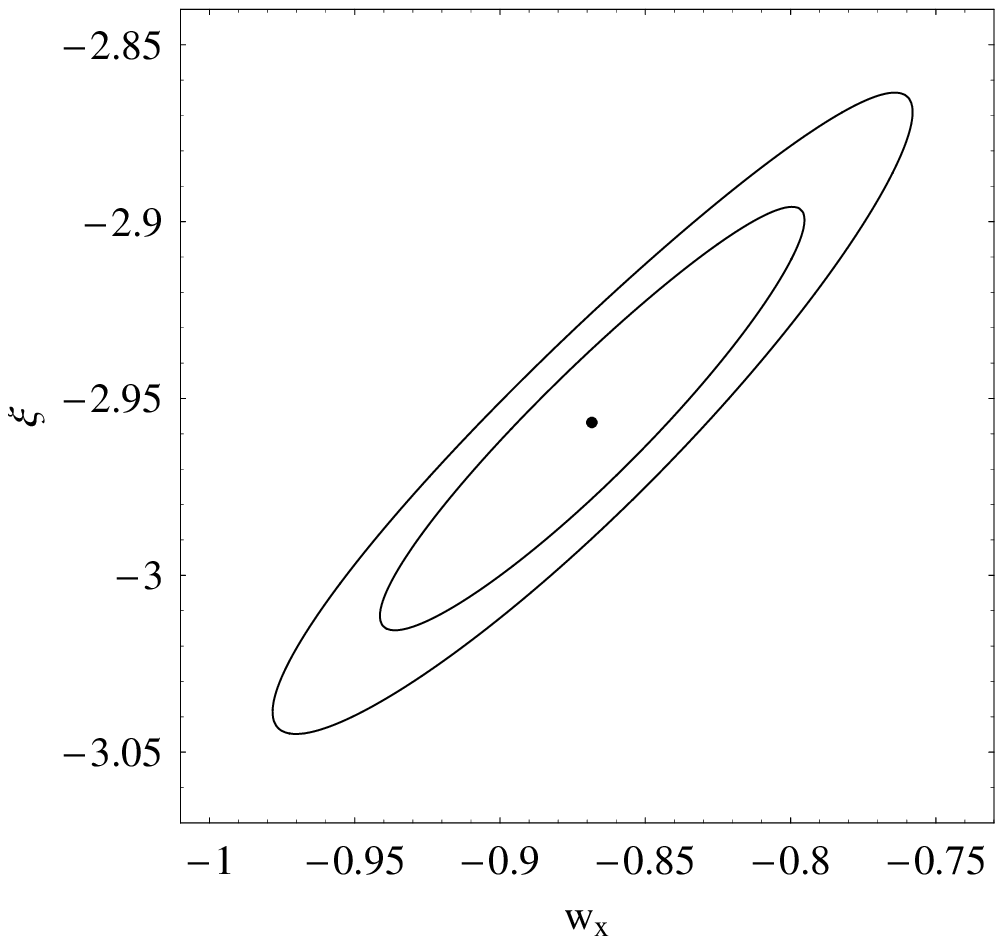}\hfill
 \includegraphics[width=0.45\textwidth]{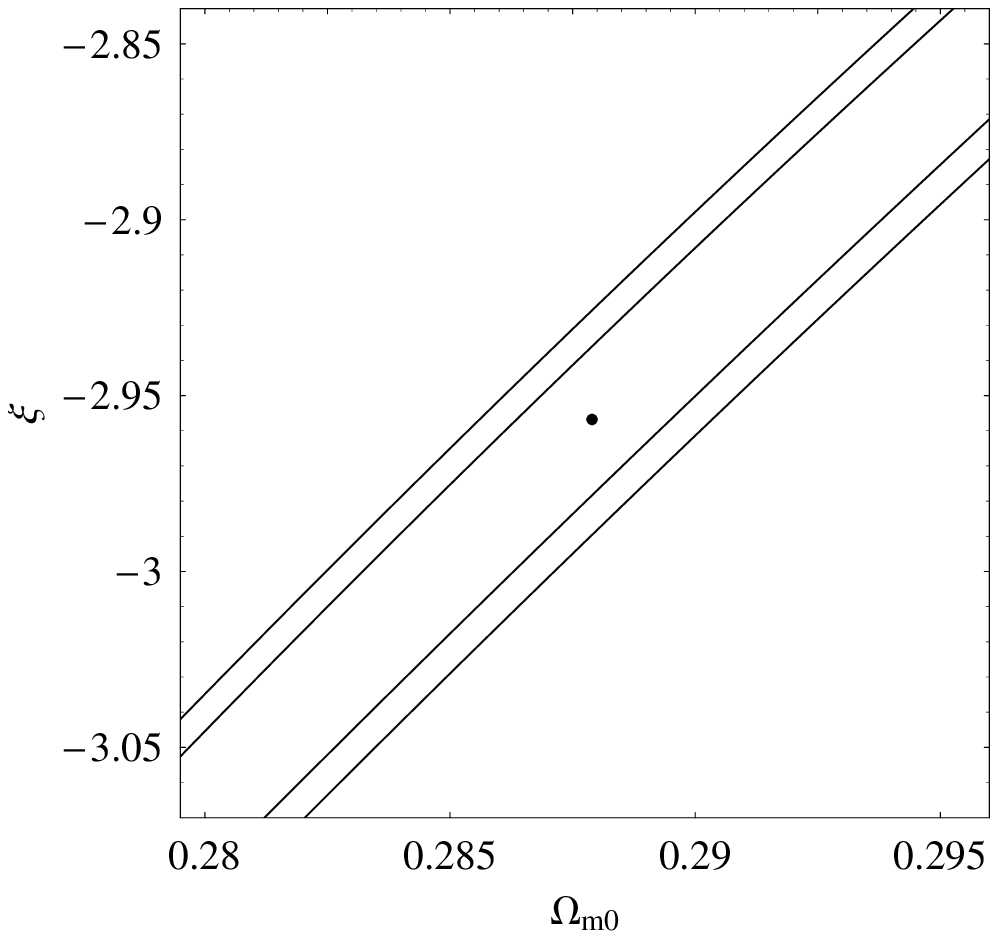}\\ \ \\
 \includegraphics[width=0.45\textwidth]{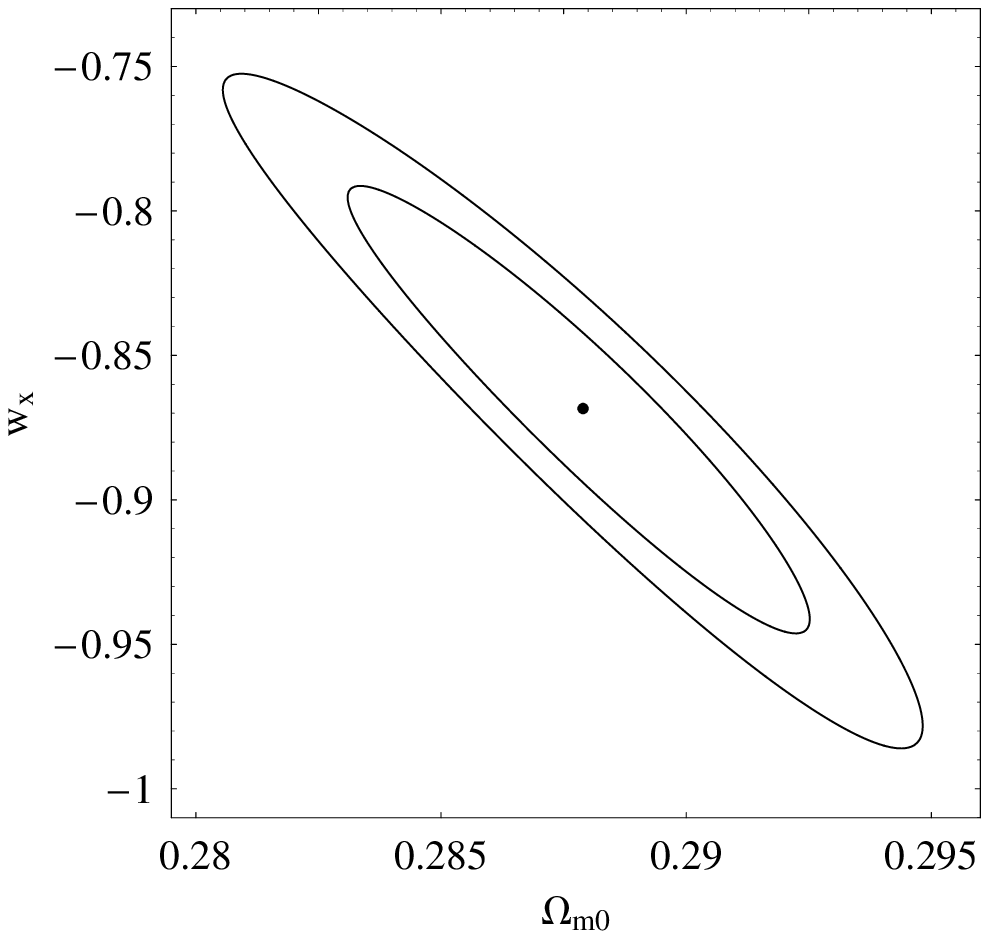}
 \caption{\label{fig3} The same as in Fig.~\ref{fig1}, except
 for the type I model characterized by $f(a)=f_0+\xi(1-a)$ without
 the condition $\xi\ge 1-\Omega_{m0}^{-1}$. See the text for
 details.}
 \end{figure}
 \end{center}

%======================================================================

\vspace{-7mm}  % used here just for a more comfortable typesetting

So, we give up the condition $\xi\ge 1-\Omega_{m0}^{-1}$ in the
 case of $f(a)=f_0+\xi(1-a)$. This means that $\rho_{_X}$ might
 be negative in the early universe. In fact,
 Guo {\it et al.}~\cite{r15} also explicitly include this
 possibility. Since such a negative energy appears in phantom
 models~\cite{r29} and modified gravity models~\cite{r30}, it
 is reasonable to consider this possibility. Without the
 condition $\xi\ge 1-\Omega_{m0}^{-1}$, by minimizing the
 corresponding total $\chi^2$ in Eq.~(\ref{eq17}), we find the
 best-fit parameters $\Omega_{m0}=0.288$, $w_{_X}=-0.868$ and
 $\xi=-2.957$, whereas $\chi^2_{min}=467.718$.
 In Fig.~\ref{fig3}, we also present the corresponding $68\%$
 and $95\%$ confidence level contours in $w_{_X}-\xi$ plane,
 $\Omega_{m0}-\xi$ plane and $\Omega_{m0}-w_{_X}$ plane.
 Obviously, these results are significantly better than the
 ones with the condition $\xi\ge 1-\Omega_{m0}^{-1}$, whereas
 the corresponding $\chi^2_{min}=467.718$ is also better.

Finally, one might consider the logarithmic case with
 \be{eq29}
 f(a)=f_0+\xi\ln a,
 \ee
 which can be regarded as a linear expansion of $f(a)$ with respect
 to the so-called $e$-folding time ${\cal N}=\ln a$ in the
 literature. This case seems attractive since in Eq.~(\ref{eq21})
 the integration is with respect to $\ln a$. However, in this case,
 $f(a)$ (and hence $\rho_{_X}$) diverges when $a\to 0$ in the
 early universe. So, we do not consider the logarithmic case in
 type~I model.

%============================= section 4 ===================================

\section{Type II models}\label{sec4}

%============================= section 4.1 ===================================

\subsection{Equations}\label{sec4a}

In this section, we turn to type~II models, which are
 characterized by Eq.~(\ref{eq7}) whereas $\epsilon(a)$ can be
 any function of $a$. Substituting Eq.~(\ref{eq7}) into
 Eq.~(\ref{eq1}), we can easily find that the corresponding
 interaction term is given by
 \be{eq30}
 Q=H\rho_m\left[\,\epsilon(a)+a\epsilon^\prime(a)\ln a\,\right].
 \ee
 It is worth noting that in type~II models, there is {\em no}
 condition like Eq.~(\ref{eq22}) in type~I models to reduce
 the number of free model parameters. If there are at least two
 parameters in $\epsilon(a)$, adding $\Omega_{m0}$ and $w_{_X}$, we
 should have four free model parameters or even more. In this case,
 the constraints will be very loose, and the calculations will
 be very involved. Instead, we follow Costa and Alcaniz~\cite{r16}
 to consider the case with a fixed $w_{_X}=-1$, namely, the role of
 dark energy is played by a decaying $\Lambda$~\cite{r12,r16}.
 Therefore, Eq.~(\ref{eq2}) becomes
 \be{eq31}
 \dot{\rho}_\Lambda=-Q
 =-H\rho_m\left[\,\epsilon(a)+a\epsilon^\prime(a)\ln a\,\right].
 \ee
 Substituting Eq.~(\ref{eq7}) into Eq.~(\ref{eq31}), we have
 \be{eq32}
 \frac{d\rho_\Lambda}{da}=-\rho_{m0}\,a^{-4+\epsilon(a)}
 \left[\,\epsilon(a)+a\epsilon^\prime(a)\ln a\,\right].
 \ee
 We can integrate Eq.~(\ref{eq32}) to obtain
 \be{eq33}
 \rho_\Lambda=\rho_{m0}\int^1_a
 \tilde{a}^{-4+\epsilon(\tilde{a})}\left[\,\epsilon(\tilde{a})
 +\tilde{a}\epsilon^\prime(\tilde{a})\ln\tilde{a}\,\right]d\tilde{a}
 +\rho_{\Lambda0}\,.
 \ee
 Substituting Eqs.~(\ref{eq33}) and~(\ref{eq7}) into Friedmann
 equation, we find that
 \be{eq34}
 E^2=\frac{H^2}{H_0^2}=
 \Omega_{m0}\,\theta(a)+\left(1-\Omega_{m0}\right),
 \ee
 where
 \be{eq35}
 \theta(a)\equiv a^{-3+\epsilon(a)}+\int^1_a
 \tilde{a}^{-4+\epsilon(\tilde{a})}\left[\,\epsilon(\tilde{a})+
 \tilde{a}\epsilon^\prime(\tilde{a})\ln\tilde{a}\,\right]
 d\tilde{a}\,.
 \ee

%============================= Fig. 4 =================================

 \begin{center}
 \begin{figure}[tbhp]
 \centering
 \includegraphics[width=0.45\textwidth]{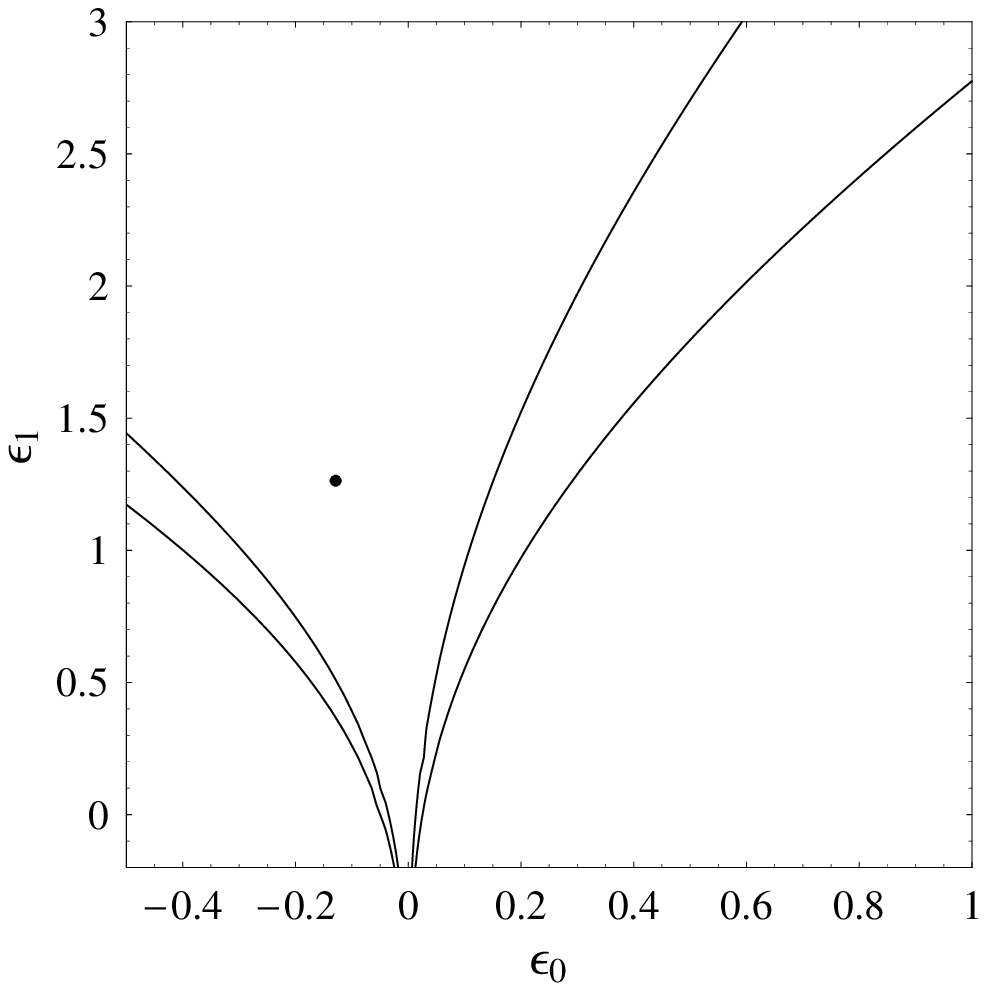}\\ \ \\
 \includegraphics[width=0.45\textwidth]{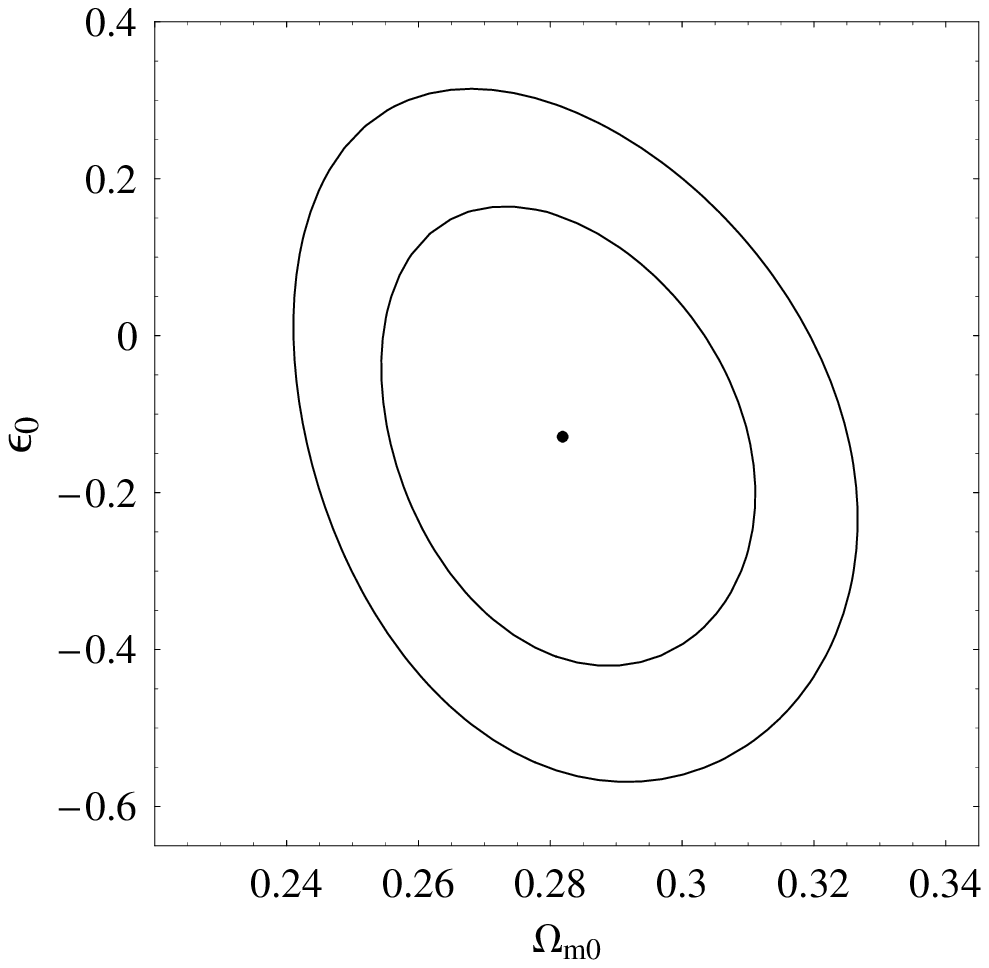}\hfill
 \includegraphics[width=0.45\textwidth]{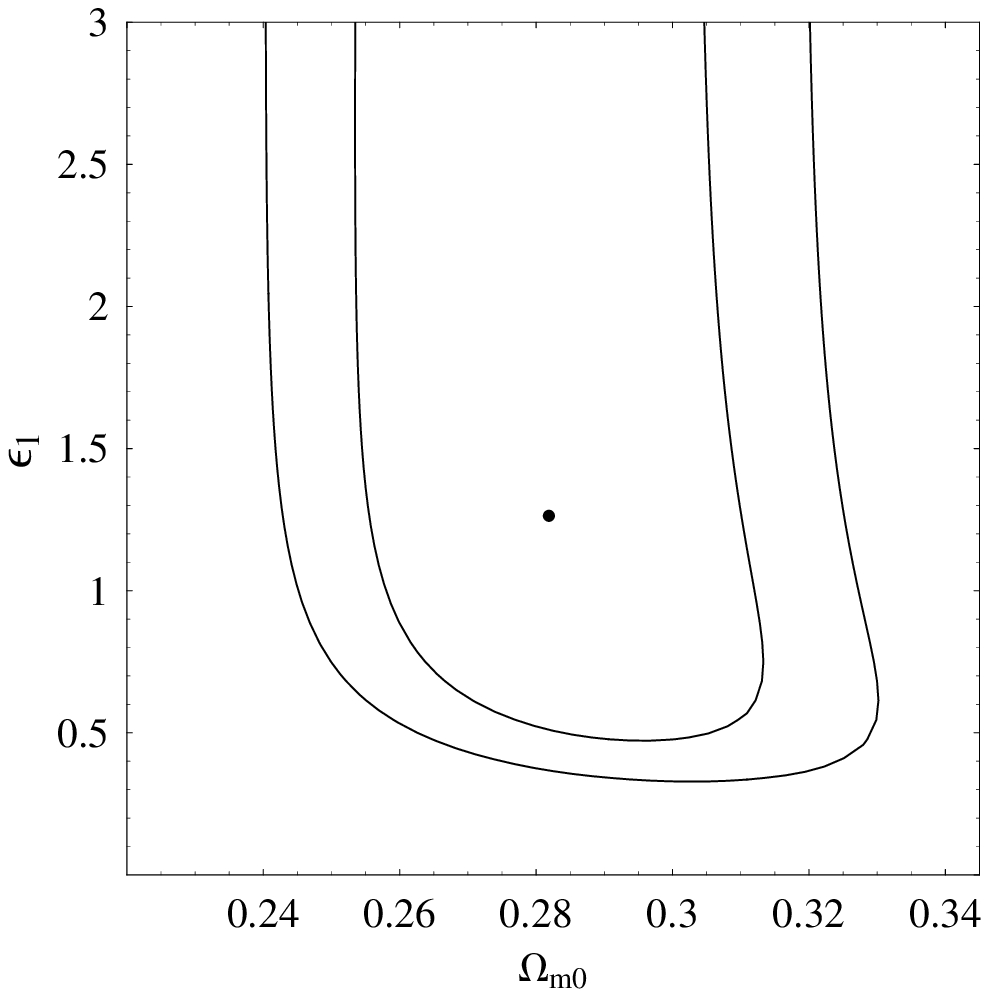}
 \caption{\label{fig4} The $68\%$ and $95\%$ confidence level
 contours in $\epsilon_0-\epsilon_1$ plane, $\Omega_{m0}-\epsilon_0$
 plane and $\Omega_{m0}-\epsilon_1$ plane for the type II model
 characterized by $\epsilon(a)=\epsilon_0\,a^{\epsilon_1}$. The
 best-fit parameters are also indicated by the black solid points.}
 \end{figure}
 \end{center}

%======================================================================

%============================= section 4.2 ===================================

\subsection{Cosmological constraints on type II models}\label{sec4b}

In this subsection, we consider the cosmological constraints
 on type II models by using the observational data given in
 Sec.~\ref{sec2}. At first, we consider the power-law case with
 \be{eq36}
 \epsilon(a)=\epsilon_0\,a^{\epsilon_1},
 \ee
 where $\epsilon_0$ and $\epsilon_1$ are constants. In this case,
 there are three free model parameters, namely, $\Omega_{m0}$,
 $\epsilon_0$ and~$\epsilon_1$. Although the cosmological
 constraints on the model characterized by Eq.~(\ref{eq36}) has
 been considered by Costa and Alcaniz~\cite{r16}, as mentioned
 in Sec.~\ref{sec1}, they have used the earlier observational data.
 Therefore, it is still worthwhile to consider the cosmological
 constraints once more in the present work by using the latest
 observational data mentioned in Sec.~\ref{sec2}. Substituting
 Eq.~(\ref{eq36}) into Eqs.~(\ref{eq34}) and~(\ref{eq35}), we
 can then fit this model to the observational data. By minimizing
 the corresponding total $\chi^2$ in Eq.~(\ref{eq17}), we find
 the best-fit parameters $\Omega_{m0}=0.282$, $\epsilon_0=-0.129$
 and $\epsilon_1=1.263$, whereas $\chi^2_{min}=465.635$. In
 Fig.~\ref{fig4}, we also present the corresponding $68\%$ and
 $95\%$ confidence level contours in $\epsilon_0-\epsilon_1$ plane,
 $\Omega_{m0}-\epsilon_0$ plane and $\Omega_{m0}-\epsilon_1$ plane.
 It is easy to see from the $\epsilon_0-\epsilon_1$ plane that
 if $\epsilon_0$ is close to zero, $\epsilon_1$ cannot be too
 negative. On the other hand, the constraint on the parameter
 $\epsilon_1$ is still very loose.

\begin{center}
 \begin{figure}[tbhp]
 \centering
 \includegraphics[width=0.45\textwidth]{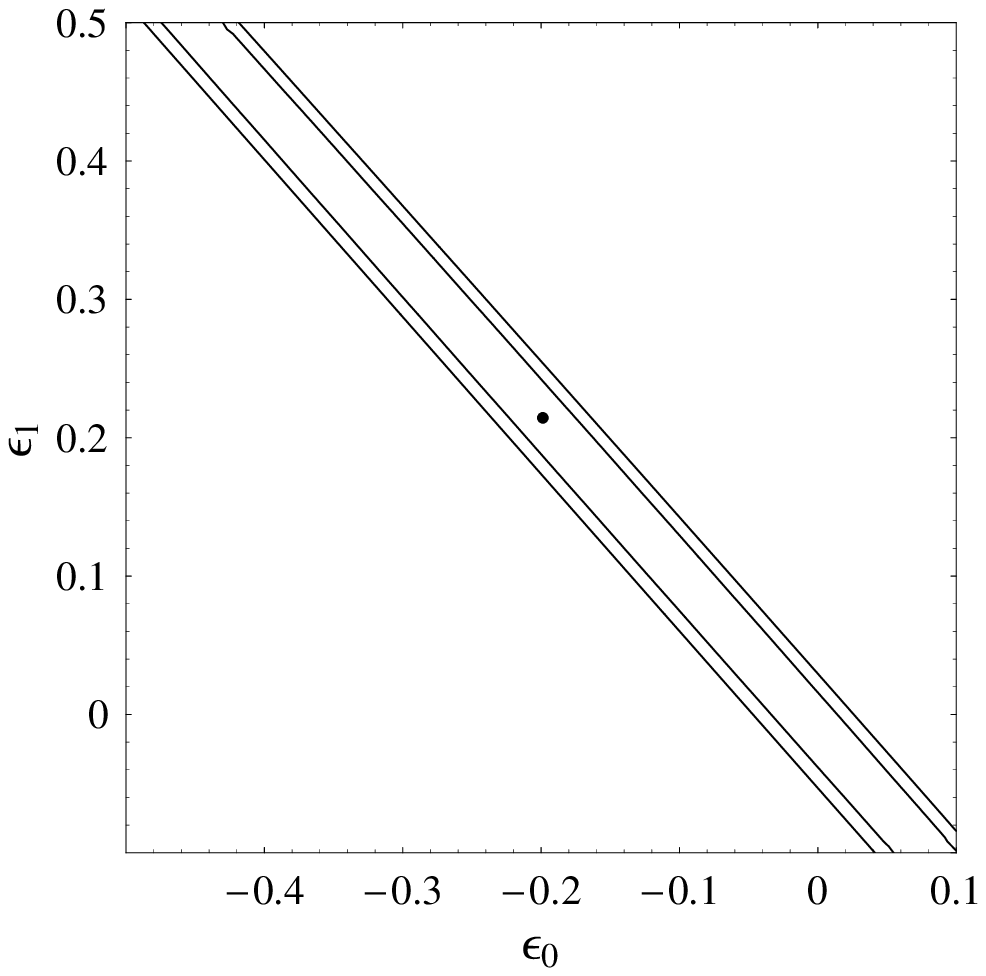}\\ \ \\
 \includegraphics[width=0.45\textwidth]{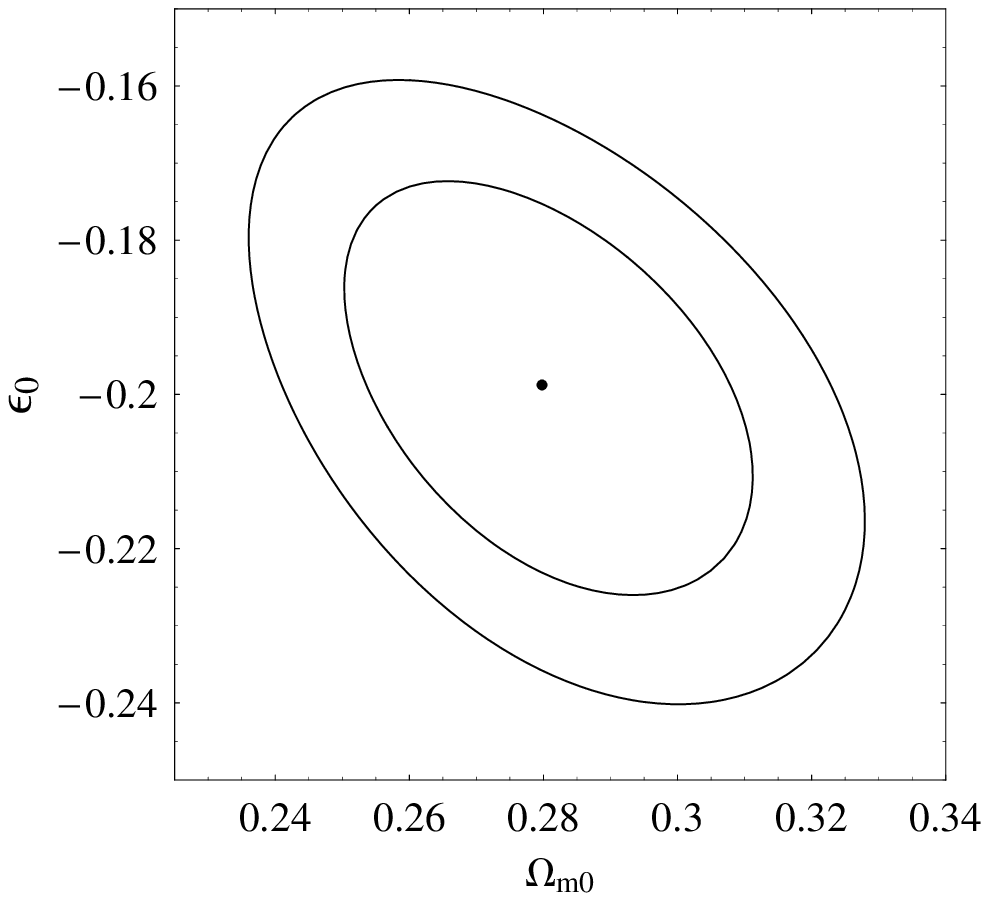}\hfill
 \includegraphics[width=0.45\textwidth]{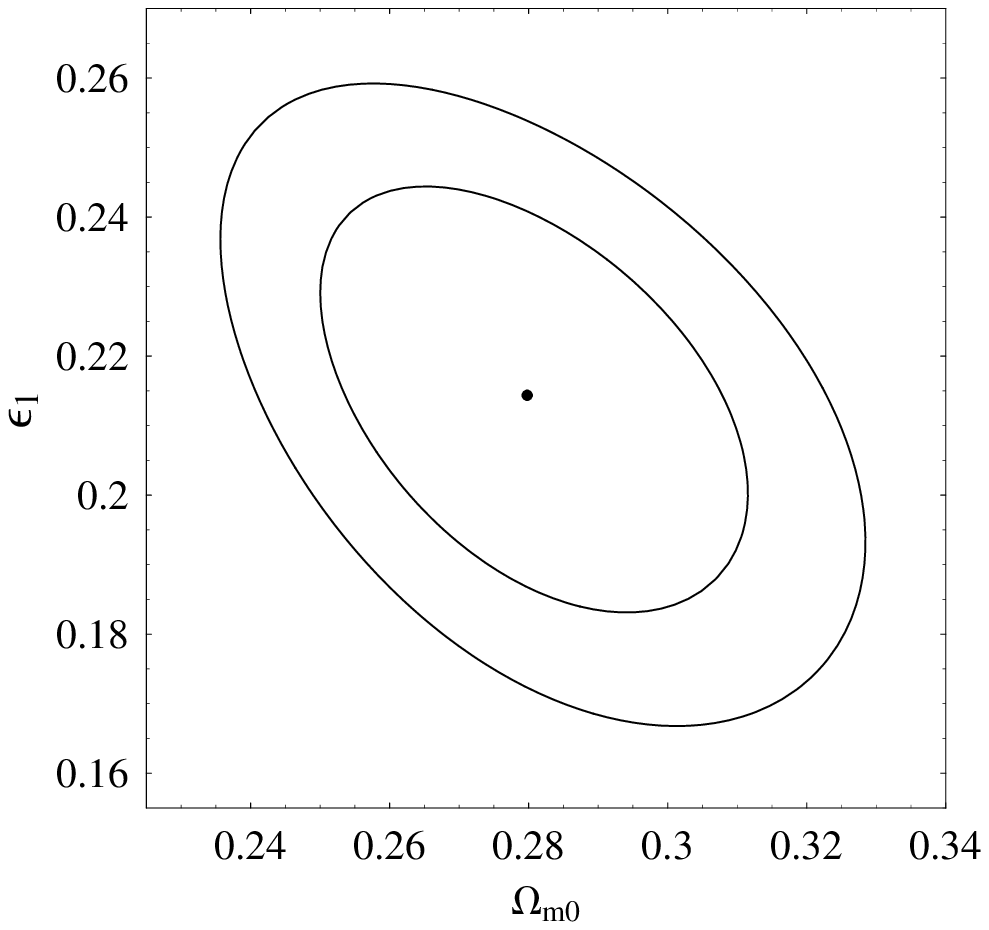}
 \caption{\label{fig5} The same as in Fig.~\ref{fig4}, except
 for the type II model characterized by
 $\epsilon(a)=\epsilon_0+\epsilon_1 (1-a)$.}
 \end{figure}
 \end{center}

Next, we turn to the CPL-like case with
 \be{eq37}
 \epsilon(a)=\epsilon_0+\epsilon_1 (1-a),
 \ee
 which can be regarded as a linear expansion of $\epsilon(a)$
 with respect to $a$, similar to the well-known CPL
 parameterization for EoS $w(a)=w_0+w_a(1-a)$~\cite{r28}.
 Substituting Eq.~(\ref{eq37}) into Eqs.~(\ref{eq34})
 and~(\ref{eq35}), we can then fit this model to
 the observational data. By minimizing the corresponding total
 $\chi^2$ in Eq.~(\ref{eq17}), we find the best-fit parameters
 $\Omega_{m0}=0.280$, $\epsilon_0=-0.199$ and $\epsilon_1=0.214$,
 whereas $\chi^2_{min}=465.604$. In Fig.~\ref{fig5}, we also
 present the corresponding $68\%$ and $95\%$ confidence level
 contours in $\epsilon_0-\epsilon_1$ plane, $\Omega_{m0}-\epsilon_0$
 plane and $\Omega_{m0}-\epsilon_1$ plane.

%============================= Fig. 5 =================================

%======================================================================

\vspace{-7mm}  % used here just for a more comfortable typesetting

\begin{center}
 \begin{figure}[tbhp]
 \centering
 \includegraphics[width=0.48\textwidth]{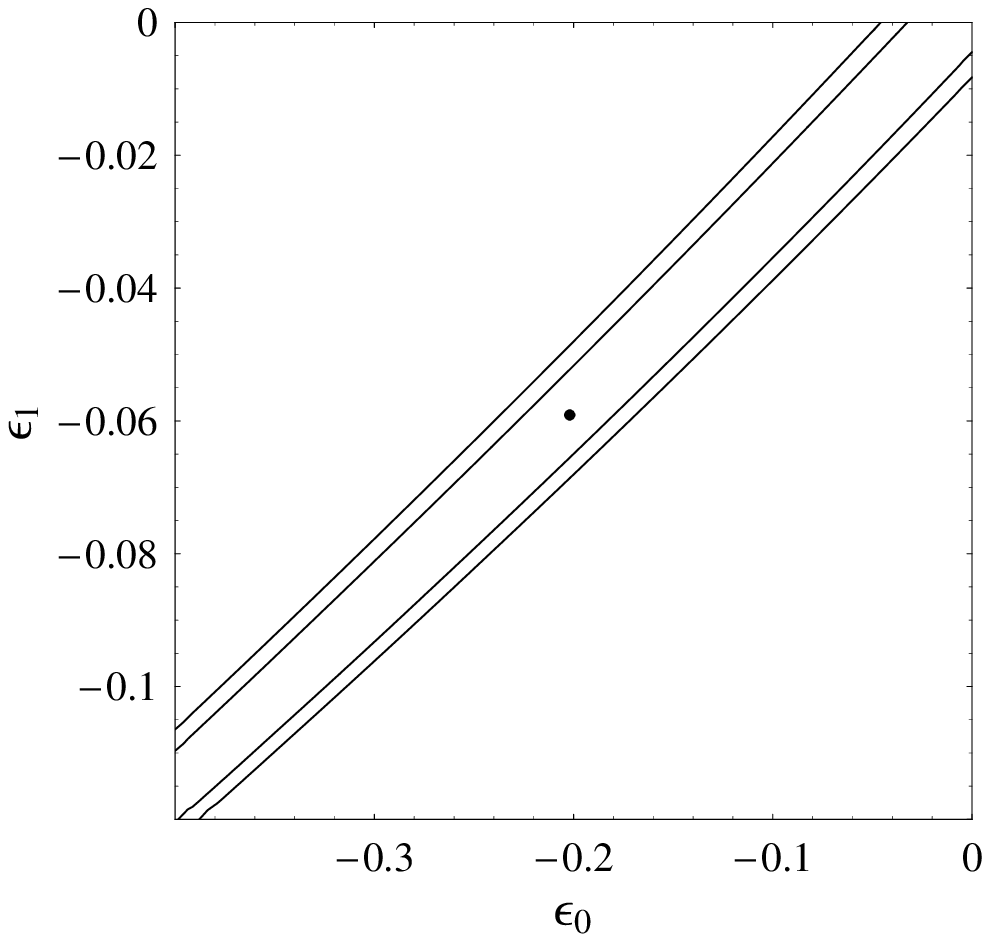}\\ \ \\
 \includegraphics[width=0.48\textwidth]{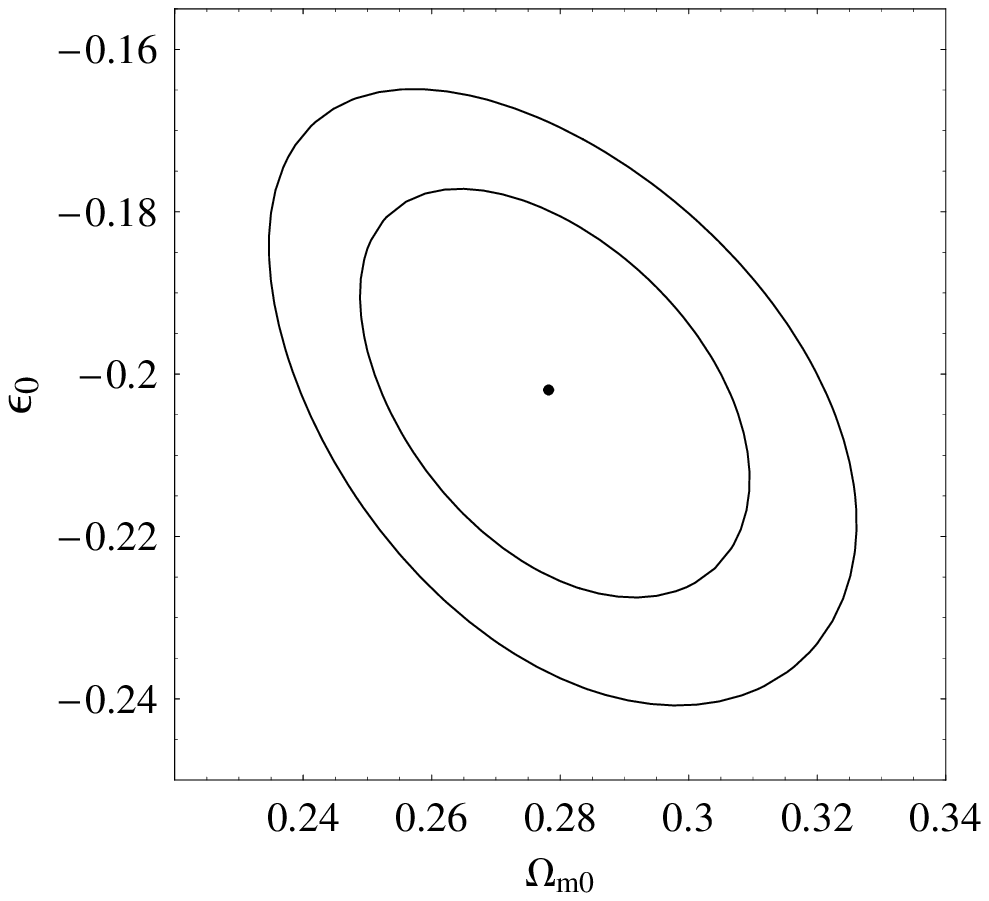}\hfill
 \includegraphics[width=0.48\textwidth]{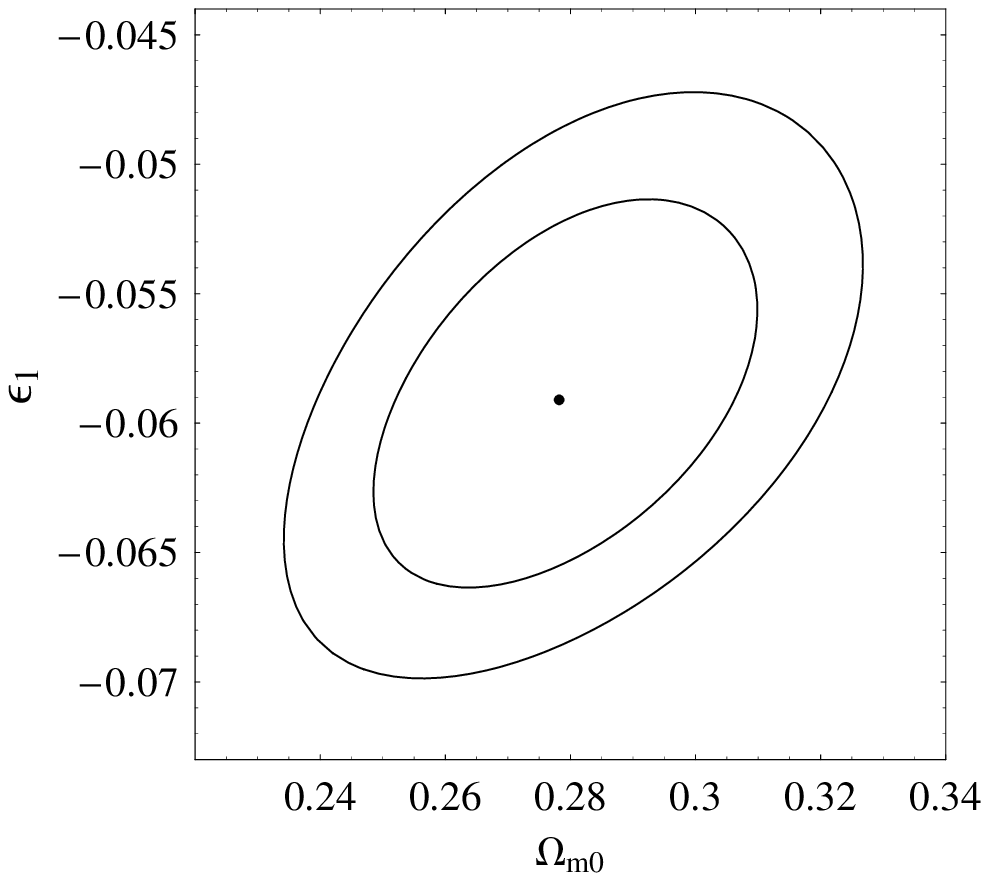}
 \caption{\label{fig6} The same as in Fig.~\ref{fig4}, except
 for the type II model characterized by
 $\epsilon(a)=\epsilon_0+\epsilon_1\ln a$.}
 \end{figure}
 \end{center}

Finally, we consider the logarithmic case with
 \be{eq38}
 \epsilon(a)=\epsilon_0+\epsilon_1\ln a\,,
 \ee
 which can be regarded as a linear expansion of $\epsilon(a)$
 with respect to the so-called $e$-folding time ${\cal N}=\ln a$ in
 the literature. Although $\epsilon(a)$ diverges when $a\to 0$
 in the early universe, unlike in the same case of type~I
 model, it does not cause any problem in type~II model, since
 $a^{-3+\epsilon(a)}\to 0^\infty \to 0$ which is regular when
 $a\to 0$. Substituting Eq.~(\ref{eq38}) into Eqs.~(\ref{eq34})
 and~(\ref{eq35}), we can then fit this model to
 the observational data. By minimizing the corresponding total
 $\chi^2$ in Eq.~(\ref{eq17}), we find the best-fit parameters
 $\Omega_{m0}=0.278$, $\epsilon_0=-0.202$ and $\epsilon_1=-0.059$,
 whereas $\chi^2_{min}=465.516$. In Fig.~\ref{fig6}, we also
 present the corresponding $68\%$ and $95\%$ confidence level
 contours in $\epsilon_0-\epsilon_1$ plane, $\Omega_{m0}-\epsilon_0$
 plane and $\Omega_{m0}-\epsilon_1$ plane.

\vspace{2mm}  % used here just for a more comfortable typesetting

%============================= Fig. 6 =================================

%======================================================================

\vspace{-7mm}  % used here just for a more comfortable typesetting

%==================== table 1 ====================

 \begin{table}[tbh]
 \begin{center}
 \begin{tabular}{llllllll}\hline\hline
 Model & $\Lambda$CDM & IPL & ICPLw
 & ICPLwo & IIPL & IICPL & IILog \\ \hline
 Best fits & $\Omega_{m0}=0.278$~ & $\Omega_{m0}=0.281$
 & $\Omega_{m0}=0.218$ & $\Omega_{m0}=0.288$ & $\Omega_{m0}=0.282$~
 & $\Omega_{m0}=0.280$~ & $\Omega_{m0}=0.278$\\
 & & $w_{_X}=-0.982$~ & $w_{_X}=-0.483$~ & $w_{_X}=-0.868$~
  & $\epsilon_0=-0.129$ & $\epsilon_0=-0.199$ & $\epsilon_0=-0.202$\\
 & & $\xi=2.988$ & $\xi=-3.584$ & $\xi=-2.957$ & $\epsilon_1=1.263$
  & $\epsilon_1=0.214$ & $\epsilon_1=-0.059$\\ \hline
 $\chi^2_{min}$ & 466.317 & 465.604 & 563.77 & 467.718
 & 465.635 & 465.604 & 465.516\\
 $k$ & 1 & 3 & 3 & 3 & 3 & 3 & 3\\
 $\chi^2_{min}/dof~$ & 1.172 & 1.176 & 1.424 & 1.181 & 1.176
 & 1.176 & 1.176\\
 $\Delta$BIC & 0 & 11.265 & 109.431 & 13.379 & 11.296 & 11.265
 & 11.177\\
 $\Delta$AIC & 0 & 3.287 & 101.453 & 5.401 & 3.318 & 3.287
 & 3.199\\
 Rank & 1 & $3\sim 4$ & 7 & 6 & 5 & $3\sim 4$ & 2\\
 \hline\hline
 \end{tabular}
 \end{center}
 \caption{\label{tab1} Summarizing all the 7 models considered
 in this work. Here, we label the type~I models characterized
 by $f(a)=f_0\,a^\xi$, $f(a)=f_0+\xi(1-a)$ with the condition
 $\xi\ge 1-\Omega_{m0}^{-1}$, and $f(a)=f_0+\xi(1-a)$ without
 the condition $\xi\ge 1-\Omega_{m0}^{-1}$ as IPL, ICPLw and
 ICPLwo, respectively. Also, we label the type~II models
 characterized by $\epsilon(a)=\epsilon_0\,a^{\epsilon_1}$,
 $\epsilon(a)=\epsilon_0+\epsilon_1 (1-a)$, and
 $\epsilon(a)=\epsilon_0+\epsilon_1\ln a$ as IIPL, IICPL and
 IILog, respectively.}
 \end{table}

%=================================================

%============================= section 5 ===================================

\section{Summary and discussions}\label{sec5}

In this work, we considered the cosmological constraints on
 the interacting dark energy models. We generalized the models
 considered previously by Guo {\it et al.}~\cite{r15}, Costa
 and Alcaniz~\cite{r16}, and we have discussed two general
 types of models: type~I models are characterized by
 $\rho_{_X}/\rho_m=f(a)$ and $f(a)$ can be any function of
 scale factor $a$, whereas type~II models are characterized
 by $\rho_m=\rho_{m0}\,a^{-3+\epsilon(a)}$ and $\epsilon(a)$
 can be any function of $a$. We obtained the cosmological
 constraints on the type~I and~II models with power-law,
 CPL-like, logarithmic $f(a)$ and $\epsilon(a)$ by using the
 latest observational data.

Some remarks are in order. Firstly, here we briefly justify
 the interaction forms considered in the present work. We take
 type~I models as examples. For the power-law case with
 $f(a)=f_0\,a^\xi$ in Eq.~(\ref{eq23}), noting that in the case
 without interaction $\rho_{_X}\propto a^{-3(1+w_{_X})}$ and
 $\rho_m\propto a^{-3}$, from definition Eq.~(\ref{eq9}), it is
 reasonable to parameterize $f(a)=\rho_{_X}/\rho_m\propto a^\xi$,
 where $\xi$ measures the severity of the coincidence
 problem~\cite{r14,r15}. For the CPL case with
 $f(a)=f_0+\xi(1-a)$ in Eq.~(\ref{eq26}) and the logarithmic
 case with $f(a)=f_0+\xi\ln a$ in Eq.~(\ref{eq29}), noting that
 the Taylor series expansion of any function $F(x)$ is given by
 $F(x)=F(x_0)+F_1\,(x-x_0)+(F_2/\,2!)\,(x-x_0)^2+
 (F_3/\,3!)\,(x-x_0)^3+\dots$, the CPL and logarithmic cases
 can be regarded as the Taylor series expansion of $f$ with
 respect to the scale factor $a$ and the $e$-folding time
 ${\cal N}=\ln a$ up to first order (linear expansion), similar
 to the well-known EoS parameterizations $w(a)=w_0+w_a(1-a)$
 and $w(z)=w_0+w_1\,z$.

Secondly, we would like to briefly consider the comparison of
 these models. For convenience, we also consider the well-known
 $\Lambda$CDM model in addition. Fitting $\Lambda$CDM model to
 the observational data considered in the present work, it is
 easy to find the corresponding best-fit parameter
 $\Omega_{m0}=0.278$, whereas $\chi^2_{min}=466.317$.
 A conventional criterion for model comparison in
 the literature is $\chi^2_{min}/dof$, in which the degree of
 freedom $dof=N-k$, whereas $N$ and $k$ are the number of data
 points and the number of free model parameters, respectively.
 We present the $\chi^2_{min}/dof$ for all the 7 models in
 Table~\ref{tab1}. On the other hand, there are other
 criterions for model comparison in the literature, such as
 Bayesian Information Criterion (BIC) and Akaike Information
 Criterion (AIC). The BIC is defined by~\cite{r33,r35}
 \be{eq39}
 {\rm BIC}=-2\ln{\cal L}_{max}+k\ln N\,,
 \ee
 where ${\cal L}_{max}$ is the maximum likelihood. In the
 Gaussian cases, $\chi^2_{min}=-2\ln{\cal L}_{max}$. So, the
 difference in BIC between two models is given by
 $\Delta{\rm BIC}=\Delta\chi^2_{min}+\Delta k \ln N$. The AIC
 is defined by~\cite{r34,r35}
 \be{eq40}
 {\rm AIC}=-2\ln{\cal L}_{max}+2k\,.
 \ee
 The difference in AIC between two models is given by
 $\Delta{\rm AIC}=\Delta\chi^2_{min}+2\Delta k$.
 In Table~\ref{tab1}, we also present the $\Delta$BIC and
 $\Delta$AIC of all the 7 models considered in this work.
 Notice that $\Lambda$CDM has been chosen to be the fiducial
 model when we calculate $\Delta$BIC and $\Delta$AIC.
 From Table~\ref{tab1}, it is easy to see that the rank of
 models is coincident in all the 3 criterions
 ($\chi^2_{min}/dof$, BIC and AIC). The $\Lambda$CDM model is
 the best one, whereas ICPLw model is the worst one. This
 result is consistent with the one obtained in e.g.~\cite{r35}.
 However, it is well known that $\Lambda$CDM model is plagued
 with the cosmological constant problem and the coincidence
 problem (see e.g.~\cite{r1}). On the other hand, as mentioned
 in the beginning of Sec.~\ref{sec1}, there are some observational
 evidences for the interaction between dark energy and dark
 matter, and the coincidence problem can be alleviated in
 the interacting dark energy models. Therefore, it is still
 worthwhile to study the interacting dark energy models.

%============================= acknowledgements ===================================

\section*{ACKNOWLEDGEMENTS}
We thank the anonymous referee for quite useful suggestions,
 which help us to improve this work. We are grateful to
 Professors Rong-Gen~Cai and Shuang~Nan~Zhang for helpful
 discussions. We also thank Minzi~Feng, as well as Xiao-Peng~Ma
 and Bo~Tang, for kind help and discussions. This paper was
 completed on the Easter Sunday of year 2010. This work was
 supported in part by NSFC under Grant No.~10905005, the
 Excellent Young Scholars Research Fund of Beijing Institute
 of Technology, and the Fundamental Research Fund of Beijing
 Institute of Technology.

\renewcommand{\baselinestretch}{1.2}

%============================= references ==================================

\end{document}